\documentclass{aa}
\usepackage{epsfig}
\usepackage{psfig}
\usepackage{graphicx}
\usepackage{txfonts}

\usepackage{lscape}

\titlerunning{X-ray observation of PSR~B2224$+$65}
\newcommand{\PSR}{PSR~B2224$+$65}
\begin{document}
\title{X-ray Emission Properties of the Old Pulsar \PSR}

\author{C. Y. Hui \and W. Becker}   
\date{Received 13 October 2006 / Accepted 13 March 2007}
\institute{Max-Planck Institut f\"ur Extraterrestrische Physik, 
          Giessenbachstrasse 1, 85741 Garching bei M\"unchen, Germany}

\abstract{Using archival Chandra data we studied the X-ray emission 
properties of \PSR\ and its environment. Albeit limited by photon 
statistics the spectral analysis suggests that the bulk of the emission from 
\PSR\ is non-thermal. Fitting a power-law model to the observed energy spectrum
yields a photon index of $\Gamma=1.58^{+0.43}_{-0.33}$.  The possible origin 
of the non-thermal pulsar emission is discussed in the context of the 
outer-gap model. We did not find any evidence for a compact nebula around 
\PSR\ though the Chandra data reveal the existence of an extended 
feature which appears to be associated with \PSR. It extends from the pulsar 
position about 2 arcmin to the north-west. Its orientation deviates by 
$\sim118^{\circ}$ from the pulsar's proper motion direction. Investigating 
its energy spectrum shows that the emission of this extended feature is much 
harder than that of the pulsar itself and is non-thermal in nature.
 
 }

\maketitle

\section{Introduction}

High energy radiation properties of rotation-powered pulsars 
are known to approximately correlate with their spin-down age. 
X-rays from young pulsars with age less than $\sim 10^{5}$ years 
are believed to be dominated by magnetospheric emission, characterized by 
a power-law spectrum and narrow pulse profiles. For middle-aged 
pulsars in the age bracket between $\sim 10^{5}-10^{6}$ years
the pulsars' X-radiation is dominated by cooling emission from
the hot stellar surface. This class of pulsars is observed to 
have composite spectra consisting of a soft thermal component, a 
harder thermal component from heated polar caps as well as 
some non-thermal emission. The thermal and non-thermal 
contributions in the emission of younger and middle-aged 
pulsars has already been well disentangled and studied by 
previous X-ray observatories (for recent reviews see e.g. Becker \& 
Aschenbach 2002; Kaspi et al.~2004 and references therein). 

Old neutron stars with ages $\gtrsim 10^{6}$ years have already 
cooled down to less than five hundred thousand degrees so that 
the thermal emission from their surface has faded from view for 
X-ray observatories. Recently, the much improved sensitivity of 
the XMM-Newton and Chandra observatories enabled a series of 
comprehensive studies to probe and identify the origin of the 
X-radiation from old rotation-powered pulsars. Until 
now, results have been presented for the pulsars B0950+08, B0823+26, 
J2043+2740 (Becker et al.~2004), B0628-28 (Becker et al.~2005), 
B0943+10 (Zhang, Sanwal \& Pavlov 2005), B1133+16 (Kargaltsev, 
Pavlov \& Garmire 2006) and for PSR B1929+10 (Becker et al.~2006).
For all old pulsars which were observed with sufficient photon 
statistics the X-ray emission turned out to be dominated by non-thermal 
emission originating in the pulsar's magnetosphere. X-ray pulses, if 
detected, often show narrow features. 

A very interesting member in this class of old pulsars is
PSR B2224+65. The proper motion of \PSR, $\mu_{\rm RA}$=144 mas/yr 
and $\mu_{\rm Dec}$=112 mas/yr, is among the highest observed by now 
(Manchester et al. 2005). The pulsar's interaction with the ISM produces a
magnificent bow shock nebula which was discovered in H$\alpha$
by Cordes et al.~(1993). Because the shape of this nebula resembles
that of a guitar the nebula got dubbed \emph{Guitar Nebula}. 
The pulsar has a period of about 0.68 s and a period 
derivative of $9.55\times 10^{-15}$ s/s, implying a spin-down luminosity
of\/ $\dot{E}=1.19\times 10^{33}$ erg/s and a spin-down age of somewhat
more than one million years. The distance to \PSR\, is not very well
constrained. The radio dispersion measure based distance in the Cordes
\& Lazio (2002) Galactic free electron density model is 1.86 kpc.
However, Chatterjee \& Cordes (2004) found by modeling the
head of the Guitar Nebula that the pulsar distance could be as close
as 1 kpc. For this work we adopt the distance of 1 kpc as suggested by 
Chatterjee \& Cordes (2004). 
The ephemeris of this pulsar as adopted from Manchester 
et al.~(2005) are summarized in Table~1.

In this paper we present a detailed X-ray study of PSR B2224+65
and its environment based on archival Chandra data. Brief results from 
this observation were already reported by Wong et al.~(2003) and Zavlin 
\& Pavlov (2004). The structure of this paper is as follows. In \S2 we
describe the observations and data analysis and in \S3 we summarize the
results which for the pulsar are discussed in the context of the 
outer-gap model.

 \begin{figure*}
 \centerline{\psfig{figure=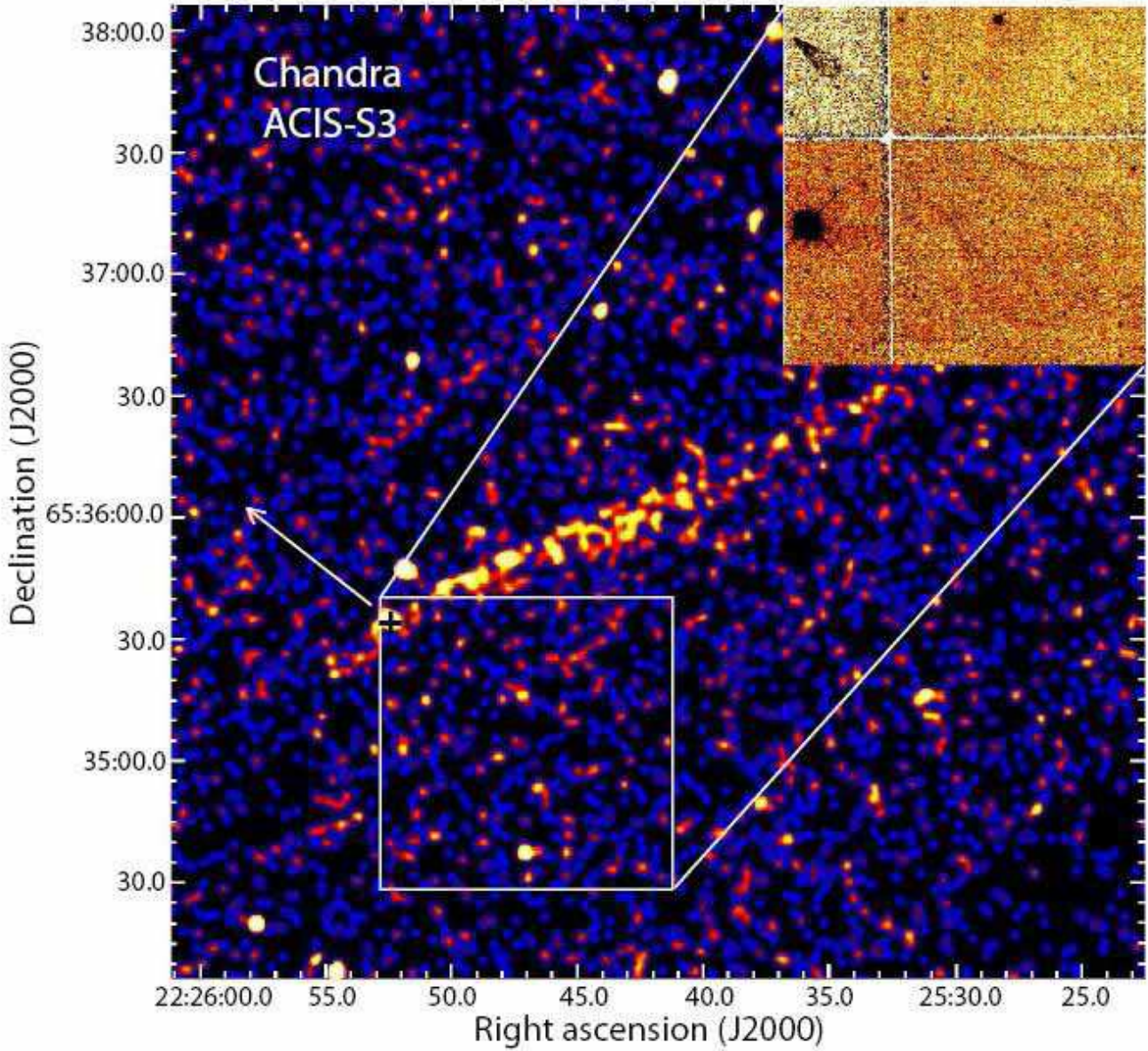,width=18cm,clip=}}
 \caption{Chandra ACIS-S3 image in the energy band $0.3-8$ keV 
 of the field around \PSR\ smoothed
 with an adaptive Gaussian filter. The black cross and white arrow 
 indicate the radio pulsar position and the pulsar's proper motion
 direction. The inset shows the H$\alpha$ image of the Guitar Nebula 
 as seen by HST WFPC2.}
 \end{figure*}

\begin{figure}
\centerline{\psfig{figure=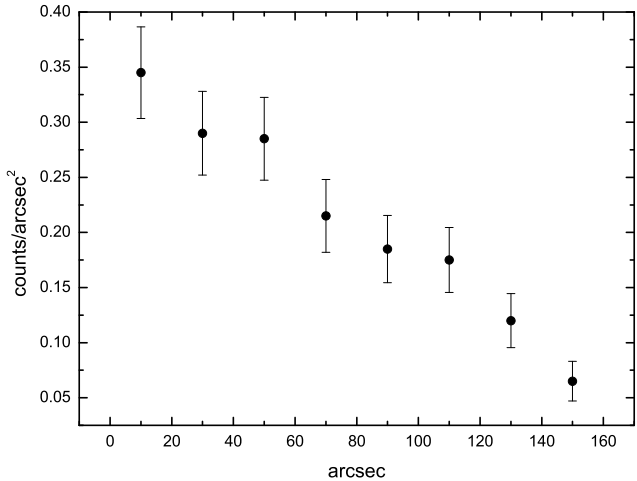,width=10cm,clip=}}
\vspace{-0.8cm}
\caption{Vignetting corrected brightness profile in the energy band $0.3-8$ keV 
of the extended feature associated with \PSR.}
\end{figure}

\begin{table}[h]
% \centering
 \caption{Proper-motion corrected ephemeris$^{a}$ of \PSR. \label{ephemeris}}
 \begin{tabular}{lc}
 \hline\hline\\
 \hline
 Right Ascension (J2000)      & $22^h 25^m 52.424^s$ \\
 Declination (J2000)          & $+65^\circ\; 35'\; 34.08"$ \\
 Pulsar Period, $P$ (s)        &  0.682538228276 \\
 Period derivative $\dot{P}$ ($10^{-15}$ s s$^{-1}$) & 9.55  \\
 Age ($10^{6}$ yrs)          & 1.13 \\
 Surface dipole magnetic field ($10^{12}$ G) & 2.58 \\
 Epoch of Period (MJD)      &  49303.00 \\
 DM (pc cm$^{-3}$)  & 36.079 \\
 DM based distance$^{b}$ (kpc)   &  1.86  \\
 Spin-down Luminosity ($10^{33}$) ergs s$^{-2}$ &  1.19\\
 \hline\\
 \end{tabular}
 $^{a}${\footnotesize Manchester et al.~(2005).}\\
 $^{b}${\footnotesize According to Cordes \& Lazio (2002).}\\
\end{table}

\section{Observations and data analysis}

\PSR\ was observed by the Chandra satellite in 2000 October 
21-22 (Obs ID: 755) with the Advanced CCD Imaging  Spectrometer 
(ACIS). The pulsar is located on the back-illuminated (BI) ACIS-S3 
chip which has a superior quantum efficiency among the spectroscopic 
array. Standard processed level-2 data were used in our study. 
The effective exposure is found to be $\sim 50$ ksec. The frame time 
of 3.2 sec does not support a timing analysis of \PSR\ to search for
X-ray pulsations. 

The Chandra image of the $4\times4$ arcmin field around \PSR\ is shown 
in Figure 1. The binning factor of the image is 0.5 arcsec. Adaptive 
smoothing with a Gaussian kernel of $\sigma<2$ pixels has been applied 
to better make visible faint diffuse emission. The X-ray point source 
is clearly seen at the radio pulsar position. The inset in Figure 1 
shows the H$\alpha$ image observed by the Wide Field Planetary Camera 2
(WFPC2) on the Hubble Space Telescope (HST). This image shows the Guitar 
Nebula. Romani et al.~(1997) reported on a $4-\sigma$ detection of diffuse 
X-ray emission associated with the H$\alpha$ nebula in ROSAT HRI data. However,
we found no evidence for any diffuse X-ray emission within the corresponding
region of the Chandra image. Rather than an extended X-ray feature which has a 
length of $\sim2$ arcmin but which deviates from the pulsar's direction
of proper motion by $\sim118^\circ$ is observed. The signal-to-noise ratio 
of this feature in the energy range 2$-$8 keV is about 6. We further compute 
the brightness profile of the feature. From the raw image with bin size of 0.5 arcsec, 
we estimate counts in eight consecutive boxes of $20\times 10$ arcsec oriented along the feature. 
The first box is centered at RA=$22^h 25^m 48.609^s$, Dec=$+65^\circ 35' 45.15"$ (J2000). The 
vignetting corrected brightness profile in the energy band $0.3-8$ keV is shown in Figure 2. 
The feature was already noticed by Wong et al.~(2003). Inspecting the radio data from the 
NRAO/VLA Sky Survey (NVSS) archive (Condon et al.~1998) did not reveal a
radio counterpart of it.

In their brief analysis, Wong et al.~(2003) claimed that the X-ray emission 
at the pulsar position is likely to be from a shocked nebula. In order to 
investigate whether there is extended nebular emission around \PSR, we have
examined the spatial nature of the emission by fitting a convolved 2-D Gaussian 
to the sub-image around the pulsar position with the point spread function (PSF) 
as the convolution kernel. The model PSF at 1.5 keV is created by using the 
CIAO tool MKPSF. The best-fitted result yields a 
FWHM of $0.55\pm0.03$ arcsec (1$-\sigma$ error) which is in agreement with 
the expected width of the Chandra PSF for a point source observed on-axis. 

For a spectral analysis of the pulsar emission we extracted events from 
a circle of 3 arcsec radius (encircled energy $\sim 99\%$) centered on 
the radio pulsar position. The spectrum from the extended X-ray feature 
was extracted from a box of $132\times15$ arcsec, oriented along the 
direction of this feature. In total, $\sim 80$ counts for the pulsar 
and $\sim 370$ counts for the feature were selected in the energy band 
$0.3-8$ keV. Even though the 
photon statistics is on the smaller size it is sufficient to provide us 
with interesting insights on the radiation emission nature of the pulsar
and the unidentified X-ray feature. 

Response files were computed by using the CIAO tools MKRMF and MKARF. 
According to the photon statistics, the spectra were dynamically binned 
so as to have at least 8 counts per bin for the pulsar and at least 30 counts 
per bin for the extended feature. The background spectrum for 
the spectral analysis of the pulsar is extracted from a 
circle near to the pulsar of 3 arcsec at 
RA=$22^{\rm h}25^{\rm m}53.893^{\rm s}$, 
Dec=$+65^{\circ}35^{'}34.79^{"}$ (J2000). For the analysis of the extended 
feature, the background spectrum is extracted from a low count region 
of $18\times15$ arcsec centered at RA=$22^{\rm h}25^{\rm m}49.113^{\rm s}$,
Dec=$+65^{\circ}36^{'}08.84^{"}$ (J2000). 
In the energy band $0.3-8$ keV, 
the net count rates for the pulsar and the 
feature are estimated to be $(1.53\pm 0.18)\times10^{-3}$ cts/s and 
$(5.59\pm 0.34)\times10^{-3}$ cts/s respectively. With the aid of PIMMS, 
the pileup fraction of the ACIS-S3 data is estimated to be $\lesssim 0.7\%$ 
which is negligible. All spectral fits were performed in the 0.3$-$8 keV 
energy range by using XSPEC 11.3.1. The degradation of the ACIS quantum 
efficiency was corrected by the XSPEC model ACISABS. The parameters of 
all fitted model spectra are summarized in Table 2. All the quoted 
errors are $1-\sigma$ and were computed for 1 parameter in interest.

For the pulsar \PSR\, we found that among the tested models a single 
power-law model describes the observed spectrum best ($\chi^{2}_{\nu}=
0.70$ for 6 D.O.F.). This model yields a column density of $N_{H}<0.9
\times10^{21}$ cm$^{-2}$, a photon index of $\Gamma=1.58^{+0.43}_{-0.33}$ 
and a normalization at 1 keV of $(2.28^{+0.75}_{-0.39})\times 10^{-6}$ 
photons keV$^{-1}$ cm$^{-2}$ s$^{-1}$. The best-fitted power-law 
spectrum and residuals are shown in Figure 3. 

The unabsorbed fluxes deduced for the best-fitting power-law model are  
$f_{X}=9.3\times 10^{-15}$ ergs cm$^{-2}$ s$^{-1}$ and 
$f_{X}=1.6\times 10^{-14}$ ergs cm$^{-2}$ s$^{-1}$ in the 0.1$-$2.4 keV and 
0.5$-$10 keV energy bands, respectively. The luminosities in these bands 
are $L_{X}$=$1.1 d^{2}_{1 {\rm kpc}}\times 10^{30}$ ergs s$^{-1}$ and
$L_{X}$=$2.0 d^{2}_{1 {\rm kpc}}\times 10^{30}$ ergs s$^{-1}$.
$d_{1 {\rm kpc}}$ denotes the distance to \PSR\ in units of 1 kpc.
The conversion efficency, $L_{X}$(0.1-2.4 keV)/$\dot{E}$, is found to be
$0.9d^{2}_{1 {\rm kpc}}\times10^{-3}$. For a distance close 
to 1 kpc as suggested by Chatterjee \& Cordes (2004), 
this is in good agreement with the relation $L_{X}$(0.1-2.4 keV)$\approx10^{-3}\dot{E}$ 
found by Becker \& Tr\"umper (1997) for rotation powered pulsars.

\begin{table*}
\caption{Spectral parameters inferred from fitting the Chandra observed 
spectrum of \PSR\ and X-ray jet. \label{spec_par}}
\begin{center}
\begin{tabular}{lccccc}
\hline\hline
Model $^{a}$ & $\chi^{2}_{\nu}$ & D.O.F. & $N_{H}$ & $\Gamma$ / $kT$ & Normalization $^{b}$ \\
      &                  &        & $10^{21}$ cm$^{-2}$ &   \\
\hline
\multicolumn{6}{c}{\PSR}\\
\hline\\
PL    &  0.70  &  6   &  $\leq 0.90$   &  $1.58^{+0.43}_{-0.33}$   & $2.28^{+0.75}_{-0.39}\times10^{-6}$   \\ 
\\
BB    & 2.07  & 6  & 0.00  & 0.46  &  $1.46\times10^{-2}$  \\ 
\\
 \hline
\multicolumn{6}{c}{Extended X-ray feature}\\
 \hline\\
PL & 0.79 & 9 & $\leq 2.25$ & $0.90^{+0.35}_{-0.24}$  & $5.66^{+3.56}_{-1.31}\times10^{-6}$  \\ 
\\
BB & 1.50 & 9 & $\leq 0.62$ & $0.91^{+0.15}_{-0.14}$ & $7.17^{+4.82}_{-2.68}\times10^{-3}$\\
\\
BREMSS & 0.86 & 9 & $1.47^{+1.44}_{-1.02}$ & $199.36^{+0.64}_{-199.36}$ & $2.46^{+0.38}_{-0.99}\times10^{-5}$ \\
\\
\hline\hline
 \end{tabular}
\end{center}
$^{a}$   {\footnotesize PL = power-law; BB = blackbody; BREMSS = Thermal bremsstrahlung}\\
$^{b}$   {\footnotesize The entry in this column depends on the model in interest. 
For the power-law model, the unit of the normalization constant is photons keV$^{-1}$ cm$^{-2}$ s$^{-1}$. 
For the blackbody model, the normalization is $(R_{km}/D_{10})^{2}$, where $R_{km}$ is the 
source radius in km and $D_{10}$ is the distance to the source in unit of 10 kpc. 
For the thermal bremsstrahlung model, the normalization constant is expressed as 
$(3.02\times10^{-15}/4\pi D^{2})\int n_{e}n_{I}dV$ where $D$ is the source distance in cm and 
$n_{e}$ and $n_{I}$ are the electron and ion densities in cm$^{-3}$.}\\
 \end{table*}

We have also considered the possibility of a purely thermal emission 
scenario. However, fitting a blackbody yields a reduced $\chi_\nu^{2}$
of 2.06 for 6 D.O.F. which invalidates this model.
Although the small number of counts does not support any fitting 
with multicomponent models, we were still able to estimate the upper 
limit for the polar cap temperature and surface temperature by adding 
a blackbody component to the best-fitting power-law model. We computed 
the confidence contours of the blackbody normalization versus the 
temperature. They are shown in Figure 4. Assuming a dipolar magnetic
field, the conventional size of a polar cap is defined by the last open field lines. 
The radius of the polar cap area is given as $r_{\rm pc}=R(2\pi R/cP)^{1/2}$, 
where $R$ is the neutron star radius, $c$ is the speed of light and $P$ 
is the rotation period of the pulsar. For \PSR, the rotation period of 
0.68 s implies a polar cap of radius $r_{\rm pc}=175$ m. With this estimate
we set a $1-\sigma$ polar cap temperature upper limit of $T^{\infty}_{\rm pc}<1.3 
\times 10^{6} K$ by assuming contribution from one polar cap only. This corresponds 
to a $1-\sigma$ upper limit on the bolometric luminosity of $L^{\rm pc}_{\rm bol}<6.5
\times 10^{29}$ ergs s$^{-1}$. 
Zavlin \& Pavlov (2004) have estimated a $1-\sigma$ upper limit on the bolometric 
luminosity of $L^{\rm pc}_{\rm bol}<9.1\times 10^{29}$ ergs s$^{-1}$  
by adopting a hydrogen polar cap model. However, such 
models have many free parameters of which most are unknown for \PSR\ (e.g.~magnetic 
inclination, viewing angle). In view of this, the applicability of this model 
is restricted and a simple blackbody model regarded to provide us with a more 
conservative upper limit. If we assume that the thermal emission is emitted 
from the whole neutron star surface of 10 km radius we compute a surface
temperature upper limit to be $T^{\infty}_{\rm s}<6.1\times 
10^{5} K$ ($3-\sigma$). This limit is of the same order as those found for
the other old X-ray detected pulsars (see Becker et al.~2004, 2005, 2006). 

\begin{figure}
\centerline{\psfig{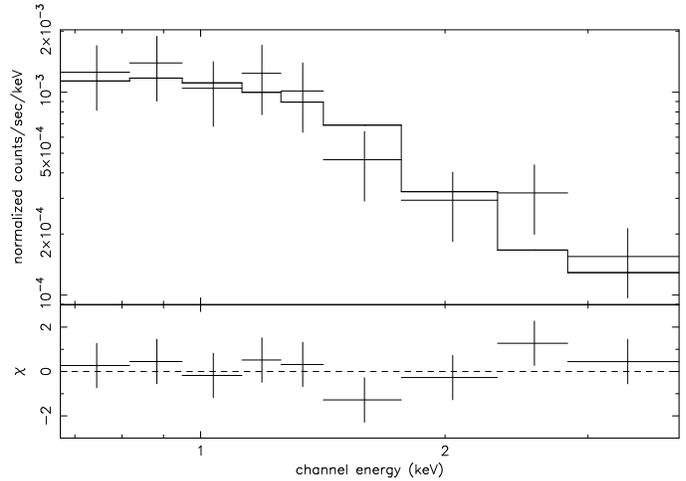}}
\caption{Energy spectrum of \PSR\, as observed with the Chandra 
 ACIS-S3 detector and fitted to an absorbed power-law model
({\it upper panel}) and contribution to the $\chi^{2}$ fit statistic
({\it lower panel}).}
\end{figure}

\begin{figure}
\centerline{\psfig{figure=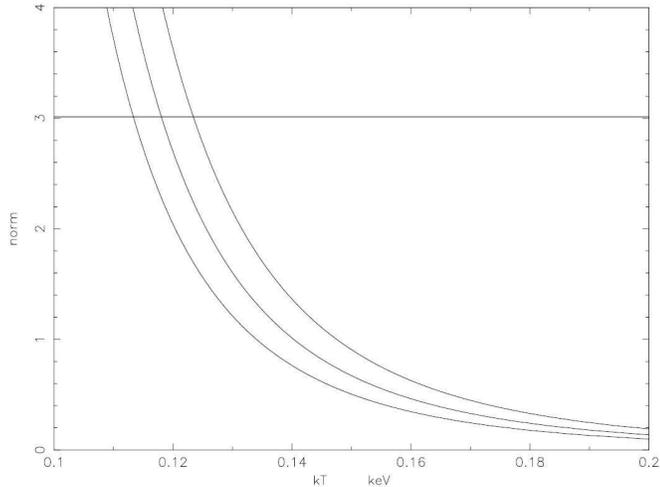,width=8.8cm,angle=0,clip=}}
\caption{Portion of the 1$\sigma$, 2$\sigma$ and 3$\sigma$ confidence contours showing the blackbody
normalization versus temperature for a blackbody model added on the best-fitted power law
model.The horizontal line at a normalization of 3.06 corresponds to a polar cap radius of
175 m and a pulsar distance of 1 kpc as suggested by Chatterjee \& Cordes (2004).}
\end{figure}

For the linearly extended feature we found that the spectrum is best-fitted by a 
single power-law model ($\chi^{2}_{\nu}=0.79$ for 9 D.O.F.). This suggests 
a non-thermal origin of its emission. The model inferred column absorption 
is found to be $N_{H}< 2.25\times 10^{21}$ cm$^{-2}$. The photon index is 
$\Gamma=0.90^{+0.35}_{-0.24}$ and the normalization at 1 keV is 
$(5.66^{+3.56}_{-1.31})\times 10^{-6}$ photons keV$^{-1}$ cm$^{-2}$ s$^{-1}$. 
The best-fitted power-law spectrum and residuals are shown in Figure 5. 
The unabsorbed fluxes of the extended feature deduced for the best-fitted 
model parameters and the energy ranges 0.1$-$2.4 keV and 0.5$-$10 keV 
are $f_{X}=2.1\times 10^{-14}$ ergs cm$^{-2}$ s$^{-1}$ and $f_{X}=1.0
\times 10^{-13}$ ergs cm$^{-2}$ s$^{-1}$, respectively. 
The corresponding luminosities in the  0.1$-$2.4 keV and 0.5$-$10 keV bands are
$L_{X}=2.5d^{2}_{1 {\rm kpc}}\times 10^{30}$ ergs s$^{-1}$ and
$L_{X}=1.2d^{2}_{1 {\rm kpc}}\times 10^{31}$ ergs s$^{-1}$, respectively. 
Fitting the spectrum of the X-ray feature with the blackbody model does not yield an acceptable 
goodness-of-fit ($\chi^{2}_{\nu}=1.50$ for 9 D.O.F.). We have also tried to fit the spectrum with a 
thermal bremsstrahlung model. However, the model implies a temperature as high 
as 200 keV and hence can be rejected simply because it is not physical. 

\begin{figure}
 \centerline{\psfig{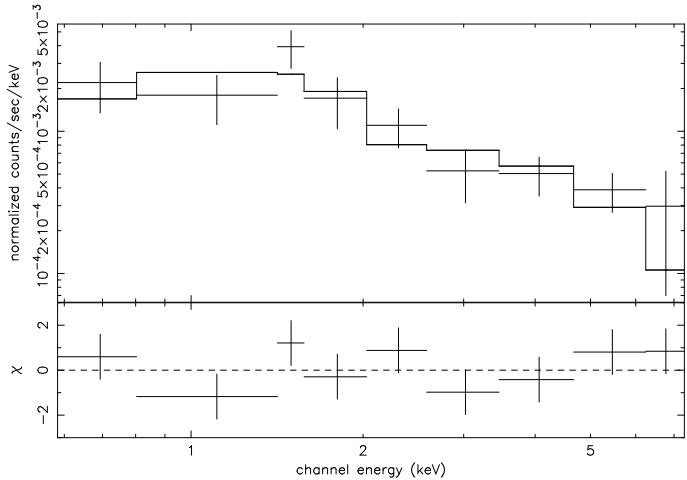}}
 \caption{Energy spectrum of extended feature detected in the
 field of \PSR, as observed with Chandra ACIS-S3 detector and 
 fitted to an absorbed power-law model ({\it upper panel}) and 
 contribution to the $\chi^{2}$ fit statistic ({\it lower panel}).}
\end{figure}

\section{Discussion \& Conclusion}
According to our analysis, the observed X-rays of \PSR\ are of non-thermal 
origin, suggesting a magnetospheric radiation dominated emission scenario. 
To model the detected X-ray spectrum with a blackbody model yields no acceptable
results. A distinct extended X-ray feature of 2 arcmin size is observed to start 
at the pulsar position. Its X-ray spectrum is very hard and no counterpart of
it is found in H$\alpha$ nor in NVSS radio data. Comparing  the energy 
spectra of \PSR\ and of the extended feature it is obvious that the X-ray 
emission of the later is much harder. We found that neither a blackbody nor 
a thermal bremsstrahlung model yields a valid description for its observed 
spectrum.  This contradicts to the findings reported by Wong et al.~(2003)
who fitted the emission from the extended feature with a blackbody.  That 
the X-ray feature is not aligned with the pulsar's proper motion does not 
a priori invalidate its association with the pulsar. Markwardt \& \"Ogelman 
(1995) reported the ROSAT PSPC observation of a highly extended jet-like 
feature associated with the Vela pulsar which also is found to be entirely 
misaligned with the pulsar's proper motion vector. The recent discovery of 
the X-ray trail associated with the millisecond pulsar PSR J2124$-$3358 is 
another example (Hui \& Becker 2006). Such asymmetry might be caused by 
e.g.~the inhomogeneities of the surrounding ISM or anisotropies in the 
pulsar wind and/or bulk flow of ambient gas (see also Gaensler, Jones \& Stappers 2002). 
%In any case, the signal-to-noise ratio of the extended 
%feature is too high to interpret it as a random structure in the ISM. 
The fact that the feature starts exactly at the position of \PSR\ 
argues against an interpretation as random ISM feature.  Deeper 
future observations though are badly needed to further study this 
unique system.

Same as \PSR\, the X-ray emission of many old pulsars are found to be 
described by power-law spectral models (Becker et al. 2004, 2005, 2006) 
suggesting that their X-ray emission is largely dominated by non-thermal 
emission processes. This is further supported by the fact that the pulse 
profiles of this class of pulsars are not sinusoidal as it would be expected
from spin-modulated thermal emission (cf.~Becker et al.~2004, 2005, 2006). 
For B1133+16 (Kargaltsev et al. 2006) and B0943+10 (Zhang et al. 2005), 
Owing to the limited photon statistics, their spectra can be fitted 
equally well by both blackbody and power-law models. While we cannot discriminate 
whether the X-ray emission are thermal or non-thermal in these cases, 
we have not taken these two pulsars in the following discussion. 

According to Cheng \& Zhang (1999; hereafter CZ99), the non-thermal X-rays 
from pulsars are synchrotron radiation of electron-positron pairs which are 
created in the strong magnetic field near the neutron star surface by curvature 
photons. These photons are emitted by charged particles on their way from the 
outer-gap to the stellar surface. The fractional size of the outer-gap, $f_{0}$, 
is defined as the ratio between the mean vertical separation of the gap boundaries 
in the plane spanned by the rotation axis and the magnetic axis to the light cylinder 
radius. CZ99 estimated the fractional gap size as $f_{0}=5.5P^{26/21}B_{12}^{-4/7}$ 
and determined the existence of the outer-gap by the criteria $f_{0}< 1$. We calculated 
the $f_{0}$ for six recently studied old pulsars and tabulated them in Table 3. Five 
out of six pulsars have $f_{0}>1$ which indicates that all but one of these old pulsars 
do not have outer-gaps in the framework of this model. Rather then, the CZ99-model 
predicts that if there is any X-ray emission from these old pulsars, it should 
be thermal. This is obviously not in agreement with the observations, showing the 
limitation of this model.

A point, however, which has not been considered in this model is that the fractional 
gap size should depend on the inclination angle $\alpha$ of the magnetic axis 
vs.~the rotational axis.
As mentioned by Zhang et al.~(2004), the active region of the outer-gap should 
begin at the null charge surface ($\Omega\cdot B=0$) which is at the radial 
distance $r_{in}$ from the star. Since the fractional size reaches a minimum 
at $r_{in}$, it is more reasonable to determine whether the gap exist by the 
criteria $f(r_{in},\alpha)<1$ rather than by $f_{0}<1$. Especially for larger 
magnetic inclination angles the null charge surface is expected to move closer 
to the star, i.e.~$f(r_{in},\alpha)$ decreases with increasing $\alpha$. Using 
Equation 36 in Zhang et al.~(2004) we calculated the variation of $f(r_{in},\alpha)$ 
with $\alpha$ for several X-ray detected old pulsars and plotted the result in Figure 6. 
Interestingly, for PSRs B0950+08, B0823+26, B0628-28 and B1929+10, 
the inclination angles deduced 
from fitting the rotating vector model to the radio polarization angle swing (Everett \& 
Weisberg 2001; Becker et al. 2005) are all in the region $f(r_{in},\alpha)<1$ 
(c.f.~Figure 6 \& Table 3). The emission 
geometry for the \PSR\ is currently unknown which prevents us
to discuss whether the outer-gap can sustain in this pulsar. 

Whether the outer-gap model is of any general meaning for the description 
of the X-ray emission from old pulsars can only be constrained by observations.
It is necessary to obtain the light curves and the spectra from a much larger 
sample than currently available. 
By now it seems rather speculative to assume that those X-ray detected old pulsars, 
by chance, all have their inclination angles in that range for which the 
outer-gap model predicts non-thermal emission.   

 \begin{table*}
 \begin{center}
 \caption{The calculated $f_{0}$ and $f(r_{in},\alpha_{\rm RVM})$ of old pulsars according 
 to CZ99 and Zhang et al. (2004) respectively.}
 \begin{tabular}{lccccc}
 \hline\hline
 PSR & $P$ (s) & $B$ ($10^{12} G$) & $\alpha_{\rm RVM}$ (degree)$^{a}$ & $f_{0}=5.5P^{26/21}B_{12}^{-4/7}$ & $f(r_{in},\alpha_{\rm RVM})$\\
 \hline
 B2224+65 & 0.68 & 2.58 & - & 1.99 & - \\
 B1929+10 & 0.23 & 0.51 & 35.97 & 1.31 & 0.62\\
 B0628-28 & 1.24 & 3.02 & 70 & 3.82 & 0.58\\
 B0950+08 & 0.25 & 0.25 & 74.6 & 2.18 & 0.23\\
 B0823+26 & 0.53 & 0.98 & 81.1 & 2.54 & 0.12\\
 J2043+2740 & 0.096 & 0.35 & - & 0.55 & - \\
 \hline
 \end{tabular}\\
 \end{center}
 $^{a}$ The magnetic inclination angles $\alpha_{\rm RVM}$ are deduced from fitting the rotating
 vector model (RVM) to the radio polarization angle swing. $\alpha_{\rm RVM}$ of B0628-28 is 
 taken from Becker et al. (2005). Others are adopted from Everett \& Weisberg (2001).
 \end{table*}

 \begin{figure*}
 \centerline{\psfig{figure=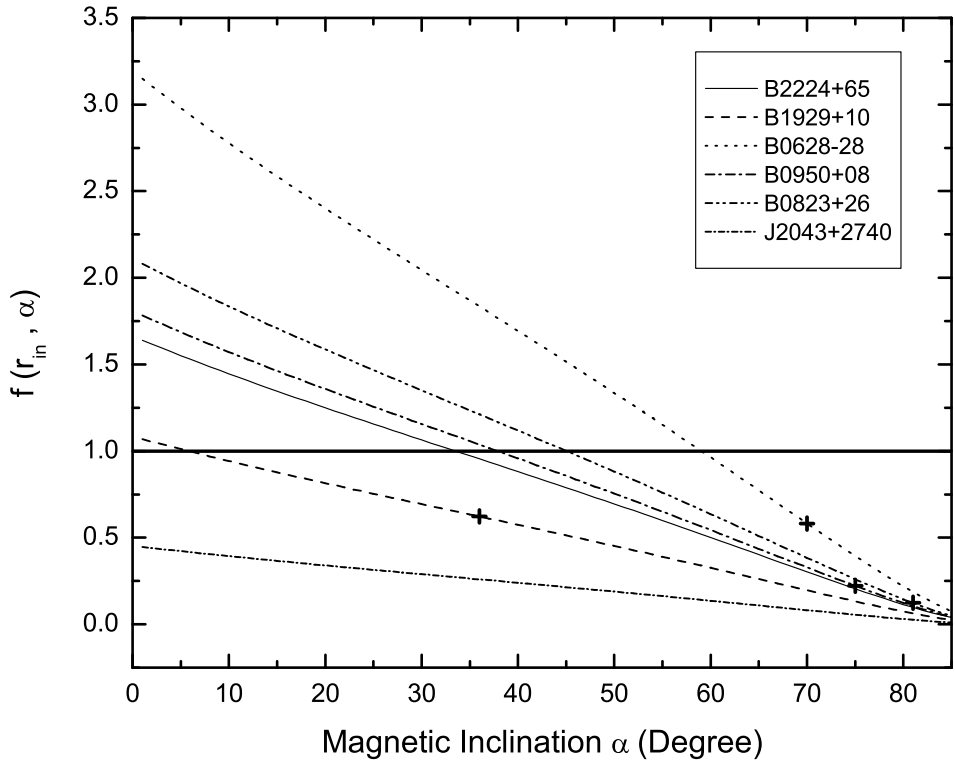,width=15cm,clip=}}
 \vspace{-0.7cm}
 \caption{Variation of the fractional size of the outer-gap at the 
 null charge surface with magnetic inclination angle for old pulsars. The crosses 
 indicate the calculated $f(r_{in},\alpha_{\rm RVM})$ for the pulsars with measured 
 magnetic inclination angle (cf. Table 3).}
  \end{figure*}

\begin{acknowledgements}
 We thank the referee David Helfand for thoroughly reading the manuscript
 and the many useful comments.
\end{acknowledgements}

\end{document}